\documentclass{webofc}
\usepackage[varg]{txfonts}

\usepackage[utf8]{inputenc}
\usepackage[english]{babel}
\usepackage{ifthen}
\usepackage{units}
\usepackage{microtype}
\usepackage[hidelinks]{hyperref}

\newcommand{\nm}{\ensuremath{N_{\mathrm{max}}}}
\newcommand{\w}{\ensuremath{\hbar\omega}}
\newcommand{\YN}{\ensuremath{Y\!N}}
\newcommand{\NN}{\ensuremath{N\!N}}
\newcommand{\NNN}{\ensuremath{N\!N\!N}}

\newcommand{\chieft}{\(\chi\)EFT}
\newcommand{\nuc}[2]{\relax\ifmmode{}^{#1}{\protect\mathrm{#2}}\else\({}^{#1}\mathrm{#2}\)\fi}
\newcommand{\hnuc}[2]{\relax\ifmmode{}_{\Lambda}^{#1}{\protect\mathrm{#2}}\else\({}_\Lambda^{#1}\mathrm{#2}\)\fi}
\newcommand{\nnlosim}{NNLO\textsubscript{sim}}
\newcommand{\LamNN}{\ensuremath{\Lambda_{\NN}}}
\newcommand{\LamYN}{\ensuremath{\Lambda_{\YN}}}
\newcommand{\Tmax}{\ensuremath{T_{\rm Lab}^{\rm max}}}
\newcommand{\luv}{\ensuremath{\Lambda_{\mathrm{UV}}}}
\newcommand{\lir}{\ensuremath{L_{\mathrm{IR}}}}
\newcommand{\hyp}{\hnuc{3}{H}}
\newcommand{\he}{\nuc{3}{He}}
\newcommand{\Gtb}{\ensuremath{\Gamma(\hyp {\to} \he \!+\! \pi^-)}}
\newcommand{\loyn}{Bonn--J\"{u}lich LO \YN{}}
\newcommand{\nnlosimstd}{\nnlosim(\LamNN = \unit[500]{MeV}, \Tmax =
  \unit[290]{MeV})}
\newcommand{\loynstd}{LO YN(\LamYN = \unit[600]{MeV})}
\newcommand{\thyp}{\ensuremath{\tau({}_\Lambda^3\mathrm{H})}}
\newcommand{\tl}{\ensuremath{\tau_\Lambda}}
\newcommand{\Esephyp}{\ensuremath{E_{\mathrm{sep}}({}_\Lambda^3\mathrm{H})}}
\newcommand{\abi}{ab~initio}

\newcommand{\Fig}[1]{figure~\ref{#1}}
\newcommand{\Tab}[1]{table~\ref{#1}}
\newcommand{\Reference}[1]{reference~\cite{#1}}
\newcommand{\Equation}[1]{equation~\eqref{#1}}
\newcommand{\Section}[1]{Sect.~\ref{#1}}

\usepackage{braket}

\usepackage{xcolor} %TMP
\newcommand{\newabbreviation}[3]{\newcounter{#1}\expandafter\newcommand\csname#1\endcsname[1][]{\ifthenelse{\equal{##1}{abreviate}}{#2}{\ifthenelse{\equal{##1}{fullname}}{#3}{\ifthenelse{\equal{##1}{explain}}{#3
          (#2)\stepcounter{#1}}{\ifthenelse{\value{#1}=0}{#3##1
            (#2##1)\stepcounter{#1}}{#2##1}}}}}}
\newabbreviation{HO}{\text{HO}}{harmonic oscillator}
\newabbreviation{NCSM}{\text{NCSM}}{no-core shell model}
\newabbreviation{YNCSM}{\text{Y-NCSM}}{hypernuclear no-core shell model}
\newabbreviation{IR}{\text{IR}}{infrared}
\newabbreviation{UV}{\text{UV}}{ultraviolet}
\newabbreviation{LEC}{\text{LEC}}{low-energy constant}
\newabbreviation{LO}{\text{LO}}{leading order}
\newabbreviation{NLO}{\text{NLO}}{next-to-leading order}
\newabbreviation{NNLO}{\text{NNLO}}{next-to-next-to-leading order}
\newabbreviation{CoM}{\text{CM}}{center-of-mass}
\newabbreviation{EFT}{EFT}{effective field theory}
\newabbreviation{chiEFT}{\(\chi\)EFT}{chiral effective field theory}
\newabbreviation{pw}{PW}{plane wave}
\newabbreviation{dw}{DW}{distorted wave}

\def\nmmax{68}
\def\fshift{-0.18cm}

\def\starextrapg{1.548}
\def\starextrapt{147\pm 17}

\begin{document}

\title{Lifetime of the hypertriton}

\author{\firstname{Daniel} \lastname{Gazda}\inst{1}\fnsep\thanks{\email{gazda@ujf.cas.cz}} \and
        \firstname{Axel} \lastname{P\'{e}rez-Obiol}\inst{2} \and
        \firstname{Avraham} \lastname{Gal}\inst{3} \and
        \firstname{Eliahu} \lastname{Friedman}\inst{3}}

\institute{Nuclear Physics Institute, 25068 \v{R}e\v{z}, Czech Republic
  \and
  Barcelona Supercomputing Center, 08034 Barcelona, Spain
  \and
  Racah Institute of Physics, The Hebrew University, Jerusalem 9190401, Israel
}

\abstract{%
  Conflicting values of the hypertriton lifetime \thyp{} were derived
  in relativistic heavy ion (RHI) collision experiments over the last
  decade. A very recent ALICE Collaboration measurement is the only
  experiment where the reported \thyp{} comes sufficiently close to
  the free-\(\Lambda\) lifetime \tl{}, as expected naively for a very
  weakly bound \(\Lambda\) in \hyp{}. We revisited theoretically this
  \hyp{} lifetime puzzle~\cite{Perez-Obiol:2020qjy}, using \hyp{} and
  \he{} wave functions computed within the \abi{} no-core shell model
  employing interactions derived from chiral effective field theory to
  calculate the two-body decay rate
  $\Gamma({}_\Lambda^3\mathrm{H}\to{}^3\mathrm{He}+\pi^-)$. We found
  significant but opposing contributions arising from $\Sigma NN$
  admixtures in \hyp{} and from $\pi^- -{}^3\mathrm{He}$ final-state
  interaction. To derive \thyp{}, we evaluated the inclusive $\pi^-$
  decay rate $\Gamma_{\pi^-}({}_\Lambda^3\mathrm{H})$ by using the
  measured branching ratio
  $\Gamma({}_\Lambda^3\mathrm{H}\to{}^3\mathrm{He}+\pi^-)/\Gamma_{\pi^-}({}_\Lambda^3\mathrm{H})$
  and added the $\pi^0$ contributions through the
  $\Delta I = \frac{1}{2}$ rule. The resulting \thyp{} varies strongly
  with the rather poorly known $\Lambda$ separation energy
  $E_{\mathrm{sep}}({}_\Lambda^3\mathrm{H})$ and it is thus possible
  to associate each one of the distinct RHI \thyp{} measurements with
  its own underlying value of
  $E_{\mathrm{sep}}({}_\Lambda^3\mathrm{H})$.}

\maketitle

\section{Introduction}
\label{sec:introduction}
Hypertriton \hyp{} is the lightest bound hypernucleus with spin-parity
\(J^\pi = \frac{1}{2}^+\) and isospin \(I = 0\). It can be
approximated as a bound state of a \(\Lambda\) hyperon and a deuteron
(\nuc{2}{H}) with tiny \(\Lambda\) separation energy of
\(\Esephyp=\unit[0.148(40)]{MeV}\)~\cite{2022EPJWC.27101006E}. The
lifetime of such a loosely bound system is expected to be comparable
to the lifetime of the free \(\Lambda\),
\(\tl = \unit[263(2)]{ps}\)~\cite{ParticleDataGroup:2020ssz}, which is
almost completely (99.7\%) governed by its nonleptonic
\(\Lambda\to N{+}\pi\) weak decay mode. Yet, the world average of
measured \hyp{} lifetime,
\(\thyp = \unit[223_{-11}^{+12}]{ps}\)~\cite{Eckert:2022dyz}, is by
\(\sim 20\%\) shorter than \tl{}. This so-called `hypertriton lifetime
puzzle' has been strengthened recently by the ALICE Collaboration's
\thyp{} value~\cite{ALICE:2022rib} which is consistent with \tl{} and
in tension with previous STAR~\cite{STAR:2017gxa} and
HypHI~\cite{Rappold:2013fic} Collaborations measurements. The latest
STAR Collaboration's measurement~\cite{STAR:2021orx} does not seem to
be conclusive since their reported \thyp{} value is consistent within
experimental uncertainties with \tl{} but its central value is 20\%
shorter. The measured \hyp{} lifetimes from recent RHI experiments
together with latest microscopic calculations are summarized in
\Tab{tab:lifetime}.
\begin{table}
  \centering
  \caption{\hyp{} lifetime values (in \unit[]{ps}) from recent
    RHI experiments and theoretical calculations.}
  \label{tab:lifetime}
  \begin{tabular}{cll}
    Experiment / Theory & \(\thyp\) & Reference \\ \hline
    Exp. & \(183_{-32}^{+42}\pm 37\)  & HypHI~\cite{Rappold:2013fic}\\
    Exp. & \(142_{-21}^{+24}\pm29\)   & STAR~\cite{STAR:2017gxa}\\
    Exp. & \(242_{-38}^{+34}\pm 17\)  & ALICE~\cite{ALICE:2019vlx}\\
    Exp. & \(221\pm 15 \pm 19\)	    & STAR~\cite{STAR:2021orx}\\
    Exp. & \(253\pm 11 \pm 6\)      & ALICE~\cite{ALICE:2022rib} \\
    Th.  & 256                      & Kamada et al.~\cite{Kamada:1997rv}\\
    Th.  & \(213\pm 5\)	            & Gal, Garcilazo~\cite{Gal:2018bvq}\\
    Th.  & see \Tab{tab:uvdep}      & P\'{e}rez-Obiol et al.~\cite{Perez-Obiol:2020qjy}\\
    Th.  & \(\approx \tl\)  & Hildenbrand, Hammer~\cite{Hildenbrand:2020kzu}
  \end{tabular}
\end{table}

\section{Method}
\label{sec:methodology}
\subsection{Hypertriton decay}
\label{sec:decay}
The main channels contributing to the \hyp{} decay are the \(\pi^-\)
and \(\pi^0\) mesonic modes
\begin{align*}
  \hyp &\to \he{+}\pi^- ,\, \nuc{2}{H}{+}p{+}\pi^-,\, p{+}p{+}n{+}\pi^-, \\
   \hyp &\to \nuc{3}{H}{+}\pi^0,\, \nuc{2}{H} {+} n{+}\pi^0,\, n{+}n{+}p{+}\pi^0
\end{align*}
due to weak-interaction \(\Lambda \to \pi + N\) and possibly
\(\Sigma \to \pi + N\) transitions, accompanied by the rare
non-mesonic modes \(\hyp \to \nuc{2}{H} {+}n,\, n{+}n{+}p\) due to
\(\Lambda N \to NN\). In this work, we calculate microscopically the
two-body \(\pi^-\) decay rate \Gtb{} and evaluate the inclusive
\(\pi^-\) decay rate \(\Gamma_{\pi^-}\) by employing the world average
experimental branching ratio
\(R_3= \Gtb{} / \Gamma_{\pi^-}(\hyp)=0.35\pm
0.04\)~\cite{Keyes:1973two}. To account for the contribution of the
\(\pi^0\) channels, we employ the empirical isospin \(\Delta I=1/2\)
rule, which relates the \(\pi^-\) and \(\pi^0\) rates by
$\Gamma_{\pi^-}=2\,\Gamma_{\pi^0}$. Finally, the non-mesonic
\(\Lambda N\to NN\) and true-absorption \(\pi NN\to NN\) contributions
to the decay rate, are incorporated through a reduction of the \hyp{}
lifetime \(\tau_\pi(\hyp) = 1 / \Gamma_\pi\) by 1.5\% and 0.8\%,
respectively~\cite{Rayet:1966fe,Golak:1998pj,Golak:1996hj,Perez-Obiol:2018oax}.
Here, \(\Gamma_\pi = 3/2 \cdot \Gtb/ R_3\) is the inclusive \hyp{}
pionic decay rate obtained using the relations above.

\subsection{Two-body \(\hyp\to\he+\pi^-\) decay rate}
\label{sec:twobody}
We follow \Reference{Kamada:1997rv} and write the two-body \(\pi^-\)
decay rate of \hyp{} as \cite{Perez-Obiol:2020qjy}
\begin{equation}
  \label{eq:rate}
  \Gamma^{\he} = \frac{3}{4\pi}
  \frac{M_{\he}\,q_\pi}{M_{\he} + \sqrt{m_\pi^2+q_\pi^2}}\,
  \sum_{m_{\hyp}}\sum_{m_{\he}} \left\lvert
    \langle\Psi_{\he}\phi_\pi\vert\hat{O}\vert\Psi_{\hyp}\rangle
  \right\rvert^2,
\end{equation}
where \(M_{\hyp}=\unit[2991]{MeV}\) and \(M_{\he}=\unit[2809]{MeV}\)
are the \hyp{} and \he{} masses, and
\(q_\pi= (2M_{\he})^{\frac{1}{2}}[M_{\hyp}
-(m_\pi^2+2M_{\he}M_{\hyp}-M_{\he}^2)^{\frac{1}{2}}]^{\frac{1}{2}}=\unit[114.4]{MeV}\)
is the pion momentum, with \(m_\pi=\unit[138.04]{MeV}\) the pion mass.
The summations run over spin projections \(m_{\he}\) and \(m_{\hyp}\)
of the initial \hyp{} and final \he{} wave functions \(\Psi\)
discussed in \Section{sec:yncsm}. The \(\phi_\pi\) is the \(\pi^-\)
wave function discussed in
\Section{sec:pions}. The operator \(\hat{O}\) due to the
weak-interaction \(\Lambda, \Sigma \to N{+}\pi\) transitions
\begin{equation}
  \label{eq:weakop}
  \frac{\hat{O}}{i\,G_Fm_\pi^2} =
  \sqrt{2}\left(
    \mathcal{A}_\Lambda+\frac{\mathcal{B}_\Lambda}{2\overline{M}_{\Lambda N}}
    \vec{\sigma}\cdot \vec{q}\right)\hat{P}_{t_{12}=0}
  +\frac{1}{\sqrt{2}}{\cal A}_{\Sigma^-}\hat{P}_{t_{12}=1}
\end{equation}
contains contributions from \(\Lambda\to \pi^- {+}p\), together with
\(\Sigma^-\to n{+}\pi^-\) and \(\Sigma^0\to p{+}\pi^- \) due to
\(\Sigma\! N\!N\) admixtures in the \hyp{} wave function. The
$\vec{\sigma}$ and $\vec{\tau}$ are the spin and isospin Pauli
matrices, \(G_F m_\pi^2 = 2.21 \times 10^{-7}\), \(\vec{q}\) is the
pion on-shell momentum, and
\(\overline{M}_{\Lambda N}= \unit[1027.3]{MeV}\) is the average
\(\Lambda-N\) mass. The parity-violating (PV)
\(\mathcal{A}_\Lambda=1.024\) and parity-conserving (PC)
\(\mathcal{B}_\Lambda=-9.431\) amplitudes were extracted from the free
\(\Lambda\) lifetime
\(\tl=\unit[263(2)]{ps}\)~\cite{ParticleDataGroup:2020ssz}, and the
PC/PV decay rates ratio~\cite{BESIII:2021ypr}. In
\Equation{eq:weakop}, we neglect the PC $\Sigma$
amplitudes~\cite{Donoghue:1992dd}, fix the PV
\(\Sigma^-\to n{+}\pi^-\) amplitude
\({\cal A}_{\Sigma^-}=1.364\)~\cite{Donoghue:1992dd} by
\(\tau_{\Sigma^-}=\unit[147.9]{ps}\), and use the \chieft{} Lagrangian
to relate the \(\Sigma^0\to p{+}\pi^-\) amplitude by
\({\cal A}_{\Sigma^0}=\frac{1}{\sqrt 2}\,{\cal A}_{\Sigma^-}\). The
projection operators \(\hat{P}_{t_{12}}\) select \hyp{} and \he{} wave
function components with a specific isospin \(t_{12}\) of the \(\NN\)
subcluster. For more details and origin of the numerical factors in
equations~\eqref{eq:rate} and \eqref{eq:weakop} see
\Reference{Perez-Obiol:2020qjy}.

\subsection{Nuclear and hypernuclear wave functions}
\label{sec:yncsm}
The initial-{} and final-state $\hyp$ and $\he$ wave functions in
\Equation{eq:rate} are computed within the \abi{} \NCSM{} approach
\cite{Barrett2013,navratil2009}, where nuclei and hypernuclei are
described as systems of \(A\) nonrelativistic particles interacting
through realistic nucleon--nucleon (\NN), 3-nucleon (\NNN) and
hyperon-nucleon (\YN) interactions. In this work we employed a version
of \NCSM{} formulated in translationally-invariant relative
Jacobi-coordinate \HO{} basis which is suitable for dealing with
few-body systems~\cite{Gazda2014, Wirth:2017bpw}. The many-body wave
function is cast as an expansion in a complete set of basis states
\begin{equation}
  \label{eq:expansion}
  \ket{\Psi^{J^\pi T}_E} = \sum_{N=0}^{\nm} \sum_\lambda c_{N\lambda}^{J^\pi T}
  \ket{N\lambda JT},
\end{equation}
where the \HO{} states \(\ket{N\lambda JT}\) are characterized by the
HO frequency \w{} and the expansion is truncated by the maximum number
\nm{} of HO excitations above the lowest configuration allowed by
Pauli principle. In \Equation{eq:expansion}, \(N\) is the total number
of HO excitations of all particles and \(J^{\pi}T\) are the total
angular momentum, parity and isospin. The quantum number \(\lambda\)
labels all additional quantum numbers and the sum over \(N\) is
restricted by parity to an even or odd sequence. The energy
eigenstates are obtained by solving the Schr\"odinger equation
\begin{equation}
  \label{eq:schr}
  \hat{H} \ket{\Psi^{J^\pi T}_E} = E(J^\pi T) \ket{\Psi^{J^\pi T}_E}
\end{equation}
with the intrinsic Hamiltonian
\begin{equation}
  \label{eq:Ham}
    H = \sum_{i=1}^A \frac{\vec{p}^{\,2}_{i}}{2m_i}
    + \sum_{1 \le i < j}^{A-1} V_{\NN,\, ij}
    + \sum_{1 \le i < j < k}^{A-1} V_{\NNN,\, ijk}
    + \sum_{i=1}^{A-1} V_{\YN,\, Ai} + \Delta M
    - H_{\mathrm{CM}}.
\end{equation}
Here, the masses $m_i$ and single-particle momenta $\vec{p}_i$ for
$i \le A-1 $ correspond to the nucleons and for $i = A$ to hyperons.
The \(H_{\mathrm{CM}}\) is the free \CoM{} Hamiltonian and the mass
term \(\Delta M = \sum_{i<A} m_i - M_0\) with $M_0$ the reference mass
of a hypernuclear system containing only nucleons and a $\Lambda$
hyperon is introduced to account for the mass difference of coupled
$\Lambda$-{} and $\Sigma$-hypernuclear states.

\begin{figure}
\centering
\includegraphics[width = 9cm, clip]{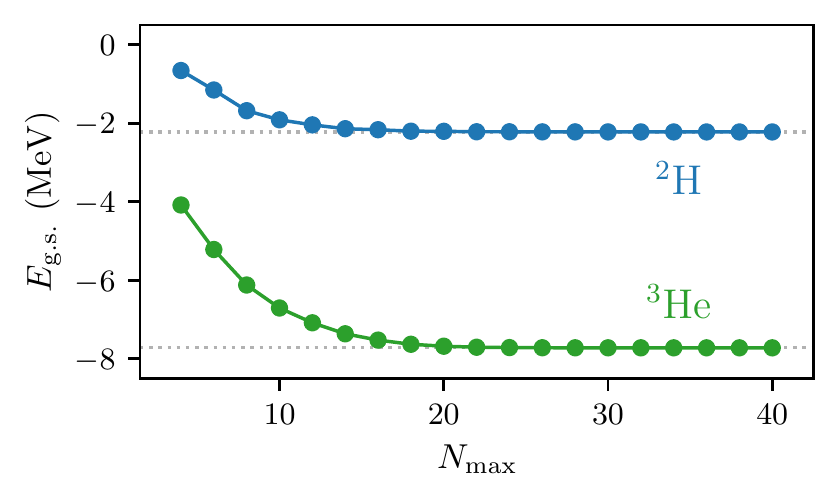}\vspace{\fshift}
\caption{Dependence of \nuc{2}{H} and \he{} ground-state energies
  \(E_{\mathrm{g.s.}}\) on the NCSM model-space truncation \nm{} for
  fixed value of the HO frequency \(\hbar\omega = \unit[14]{MeV}\),
  calculated using the \nnlosimstd{} interaction. Experimental values
  are indicated by the gray dotted lines.}
\label{fig:E2H3He}
\end{figure}
For the nuclear \(V_{\NN}{+}V_{\NNN}\) interactions in
\Equation{eq:Ham} we employ the \nnlosim{}
potentials~\cite{Carlsson:2015vda} that are based on \chiEFT{} up to
\NNLO{}. The \nnlosim{} is a family of 42 different interactions where
each potential is associated with one of seven different regulator
cutoffs $\LamNN = 450, 475, \ldots, 575, \unit[600]{MeV}$ and six
different maximum scattering energies in the laboratory system
$\Tmax = 125, 158, \ldots, 257, \unit[290]{MeV}$ at which the
experimental \NN{} scattering cross sections data base used to
constrain the respective interaction was truncated. For the purpose of
this work precise wave functions of \he{} and the `core nucleus'
\nuc{2}{H} are required. Since certain low-energy properties of
\nuc{2}{H} and \nuc{3}{He} were included in the pool of fit data,
their energies are accurately described for all \nnlosim{}
interactions,
\(E_{\nuc{2}{H}, \nuc{3}{He}} = -2.224^{(+0)}_{(-1)},
\unit[-7.717_{(-21)}^{(+17)}]{MeV}\)~\cite{Carlsson:2015vda}. For
\nnlosim{}, the NCSM ground-state (g.s.) energies of \nuc{2}{H} and
\nuc{3}{He} exhibit a good convergence with the \NCSM{} model-space
truncation \nm{} as demonstrated in \Fig{fig:E2H3He} for
\(\w=\unit[14]{MeV}\) using the \nnlosimstd{} \(\NN{}{+}\NNN{}\)
interaction. The g.s.\ energies are converged within few~\unit[]{keV}
already at \(\nm \approx 30\) for a wide range of HO frequencies.

\begin{figure}
\centering
\includegraphics[width = 9cm, clip]{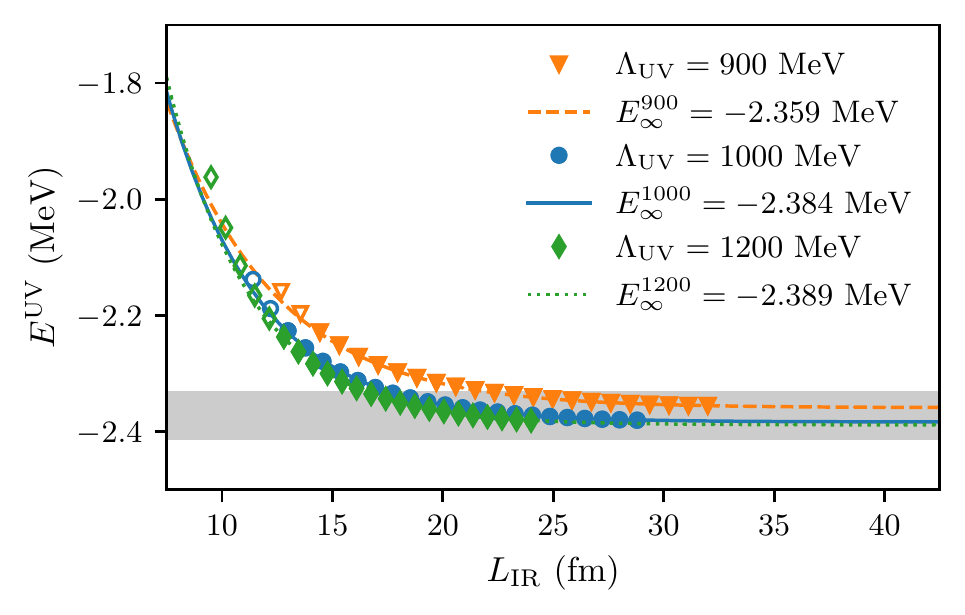}\vspace{\fshift}
\caption{The $\hyp$ ground-state energies \(E^{\mathrm{UV}}\) as a
  function of the IR length \lir{} (up to $\nm=\nmmax$) for the
  \nnlosimstd{} and \loynstd{} interactions, together with their
  extrapolated values for several fixed values of the UV cutoff,
  \(\luv{} = 900, 1000, \unit[1200]{MeV}\). Only the points marked by
  filled symbols, corresponding to particle-stable \hyp{}
  configurations, are included in the fits. Experimental
  value~\cite{2022EPJWC.27101006E} is indicated by the gray shaded
  area.}
\label{fig:EL3H}
\end{figure}

For the \YN{} sector (\(V_{\YN}\) in \Equation{eq:Ham}) we use the
\LO{} coupled-channel Bonn--J\"{u}lich SU(3)-based \chiEFT{}
model~\cite{Polinder:2006zh}. The potential consists of pseudoscalar
$\pi$, $K$, $\eta$ meson exchanges and baryon--baryon contact
interaction terms. The interaction is regularized by a smooth momentum
cutoff \(\LamYN\) ranging from 550 to \unit[700]{MeV}. Unfortunately,
calculations of \hnuc{3}{H} g.s.\ energy exhibit a slower convergence
with the \NCSM{} model-space truncation \nm{}. This can be attributed
to the very small binding energy of \hyp{} and, consequently, the long
tail of the \hyp{} wave function in coordinate space. Nevertheless,
well-converged results for the \hyp{} g.s.\ energy can be obtained for
\(\nm \approx 70\)~\cite{Htun:2021jnu}. However, as shown
in \Section{sec:results}, it is not possible to achieve full
convergence for the transition operator matrix element in
\Equation{eq:rate} even in the largest feasible \NCSM{} model spaces
and it is necessary to extrapolate the finite-space results into
infinite model space. We employ the \IR{} extrapolation scheme
developed for nuclear NCSM in~\cite{Wendt2015} and generalized
recently to hypernuclear \NCSM{} in~\cite{Gazda:2022fte}. In this
scheme, the truncation of the HO basis in terms of \nm{} and \w{} is
translated into the associated infrared (\lir{}) and ultraviolet~(UV)
(\luv{}) scales and \IR{} correction formulae can be systematically
derived for observables to extrapolate to infinite model space,
\(\lir \to \infty\). The LO correction for energies and the expected
magnitude of subleading corrections
\(\sigma_{\mathrm{IR}}\)~\cite{Forssen:2017wei} are
\begin{equation}
  \label{eq:ir}
  E^{\mathrm{UV}}(\lir) = E_\infty^{\mathrm{UV}} + a_0^\mathrm{UV}\,
  e^{-2k_\infty^{\mathrm{UV}}\lir},\quad \sigma_{\mathrm{IR}} \propto \frac{e^{-2k_\infty^{\mathrm{UV}}\lir}}{k_\infty^{\mathrm{UV}}\lir},
\end{equation}
where
\((E_\infty^{\mathrm{UV}}, a_0^{\mathrm{UV}},
k_\infty^{\mathrm{UV}})\) are parameters determined from fit to the
NCSM-calculated energies with weights proportional to the inverse of
\(\sigma_{\mathrm{IR}}\). Note that this procedure slightly differs
and should improve the one employed in~\cite{Perez-Obiol:2020qjy}.
Additional \UV{} corrections to \Equation{eq:ir} can be substantial
unless \(\luv \gg \LamNN, \LamYN\). A large-enough \luv{} scale can be
identified by performing calculations at a fixed \luv{}---by choosing
appropriate (\nm{}, \w{}) model-space parameters---and monitoring the
\UV{} dependence ~\cite{Forssen:2017wei}. We find that
\(1000 \lesssim \luv \lesssim \unit[1200]{MeV}\) is sufficient to
achieve \UV{}-convergence and to perform \IR{} extrapolations. This is
shown in \Fig{fig:EL3H} where the \hyp{} g.s.\ energy
\(E_{\hyp}^{\rm UV}\) is shown as a function of the \IR{} length
\lir{} for several fixed values of the UV scale
\(900 \leq \luv \leq \unit[1200]{MeV}\). The NCSM calculations were
performed with model space truncation up to \(\nm = \nmmax\) using the
\nnlosimstd{} \NN{} and \loynstd{} interactions. The extrapolated
g.s.\ energies \(E_{\infty}^{\rm UV}\) show only a small dependence on
the \luv{} scale of the HO basis once
\(\luv \gtrsim \unit[1000]{MeV}\). E.g., for
\(\luv=\unit[1000(1200)]{MeV}\), which translates into
\(\w=\unit[7.299(10.510)]{MeV}\) for \(\nm=68\), the finite-space
result is \(E_{\hyp}^{1000(1200)}= \unit[-2.3807(-2.3814)]{MeV}\),
while the extrapolated infinite-space result estimated using
\Equation{eq:ir} is
\(E_\infty^{1000(1200)} = \unit[-2.384(-2.389)]{MeV}\).

\subsection{Pion wave function}\label{sec:pions}
The pion wave function in \Equation{eq:rate} was generated from a
standard optical potential constrained by data of pionic atoms across
the periodic table~\cite{Friedman:2019zhc,Friedman:2020gsf}. In order
to extrapolate from the near-threshold region relevant for pionic
atoms to \(q_\pi=\unit[114.4]{MeV}\) in the \(\pi^--\he{}\) \CoM{}
system, the potential parameters were adjusted using the
\(\pi N\)~\cite{Arndt:2006bf}, as well as \(\pi\)--nucleus elastic
scattering amplitudes~\cite{Friedman:2004jh,Friedman:2005pt}.

\section{Results}
\label{sec:results}
\begin{figure}
\centering
\includegraphics[width = 9cm, clip]{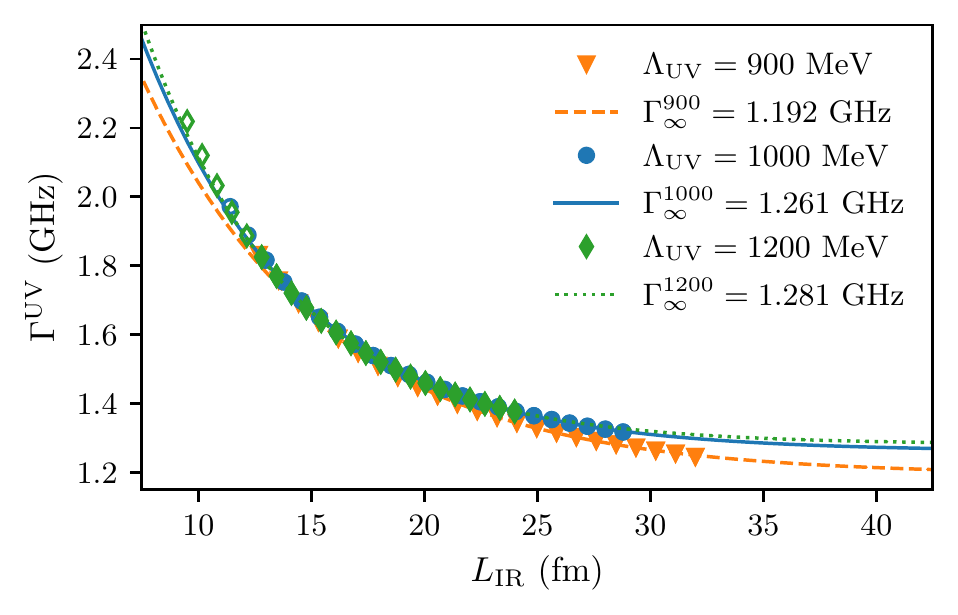}\vspace{\fshift}
\caption{The \hyp{} two-body \(\pi^-\) decay rates
  \(\Gamma^{\mathrm{UV}}\) as a function of the IR length \lir{} (up
  to $\nm=\nmmax$) for the \nnlosimstd{} and \loynstd{} interactions,
  together with their extrapolated values for several fixed values of
  the \UV{} cutoff, \(\luv{} = 900, 1000, \unit[1200]{MeV}\). The
  calculated rates include contributions of pion DW as well as
  $\Sigma NN$ channels. Marked by filled symbols are the points
  included in the fits which correspond to particle-stable \hyp{}
  configurations. The \he{} wave function was held fixed, calculated
  using \(\nm=36\) and \(\w=\unit[14]{MeV}\) NCSM model space
  parameters corresponding to a well-converged configuration.}
\label{fig:2brate}
\end{figure}

In \Fig{fig:2brate}, we show the 2-body \hyp{} \(\pi^-\) decay rate
\Gtb{} as a function of the HO \IR{} length scale \lir{} (open and
filled symbols), together with their extrapolations (solid lines), for
several values of the HO basis \UV{} cutoff \luv{}. Marked by the
solid symbols are the points included in the fit which correspond to
particle-stable \hyp{} configurations with
\(E(\hyp) < E(\nuc{2}{H})\). The \IR{} extrapolations were performed
by adapting the relations in \eqref{eq:ir} as
\(\Gamma^{\mathrm{UV}}(\lir) = \Gamma_\infty^{\mathrm{UV}}+
a_0^{\mathrm{UV}}\,\exp{(-2k_\infty^{\mathrm{UV}}\lir})\) with
\((\Gamma_\infty^{\mathrm{UV}}, a_0^{\mathrm{UV}},
k_\infty^{\mathrm{UV}})\) the fit parameters. This is motivated by
expanding the rate in powers of \([\Esephyp]^{\frac{1}{2}}\), see
below. The extrapolated values \(\Gamma_\infty^{\mathrm{UV}}\) show
only a weak dependence on the \UV{} scale for
\(\luv \gtrsim \unit[1000]{MeV}\). The rates in \Fig{fig:2brate} are
calculated using the \nnlosimstd{} and \loynstd{} interactions and
include the effects of \(\pi^-\) DW and \(\Sigma\to N{+}\pi\)
transitions due to the \(\Sigma NN\) component of the \hyp{} wave
function. Replacing the \(\pi^-\) PW by \(\pi^--\he\) DW in
\Equation{eq:rate} leads to increase of \Gtb{} by \(\approx 15\%\). On
the contrary, including the contributions of \(\Sigma\to N{+}\pi\)
transitions was found to decrease \Gtb{} by \(\approx 10\%\).

The dependence of the extrapolated \(\Lambda\) separation energies
\(E_\infty^{\mathrm{UV}}\) (\Fig{fig:EL3H}) and 2-body decay rates
\(\Gamma_\infty^{\mathrm{UV}}\) (\Fig{fig:2brate}) on the \UV{} scale
of the HO basis \luv{} can be exploited to study the relationship of
\Esephyp{} and \Gtb{}. While the \UV{} convergence for
\(\luv \lesssim \unit[1000]{MeV}\) was not fully achieved, the
calculations still give meaningful, sufficiently \IR{}-converged
results. The missing \UV{} corrections to \Equation{eq:ir} depend only
on short-range details of the employed interactions truncated by
\luv{}. In \Tab{tab:uvdep}, we show the extrapolated \(\Lambda\)
separation energies, two-body \(\pi^-\) decay rates, and lifetime
values for several HO basis \UV{} scales,
\(800 \le \luv{} \leq \unit[1400]{MeV}\). The \thyp{} values were
deduced from \Gtb{} using the procedure detailed in
\Section{sec:decay}, tacitly assuming that the branching ratio \(R_3\)
is independent of \Esephyp{}. The \(\Lambda\) separation energy
rapidly decreases with reducing the \UV{} scale from
\(\luv = \unit[1200]{MeV}\) to \(\luv = \unit[800]{MeV}\) which lowers
the two-body \(\pi^-\) rate and enhances \thyp{} due to the vanishing
overlap between the wave functions of the loosely bound \hyp{} and
\he{}. The small increase in \(E_{\mathrm{sep}}^{\mathrm{UV}}\) for
\(\luv = \unit[1400]{MeV}\) indicates that higher-order \IR{}
corrections to \Equation{eq:ir} become relevant for
\(\luv \gtrsim \unit[1400]{MeV}\). The last column of \Tab{tab:uvdep}
contains estimated values of \Gtb{} and \thyp{} corresponding to the
STAR Collaboration's reported value of
\(\Esephyp = \unit[0.41(12)(11)]{MeV}\)~\cite{STAR:2019wjm}. The
estimate was obtained by expanding \Gtb{} in powers of
\([\Esephyp]^{\frac{1}{2}}\) as
\(a\,[\Esephyp]^{\frac{1}{2}}+b\,\Esephyp\) and fixing the two
expansion parameters \((a,b)\) by fitting to the
\(E_{\mathrm{sep}}^{\mathrm{UV}}\) and
\(\Gamma_{\infty}^{\mathrm{UV}}\) from \Tab{tab:uvdep} for
\(800 \le \luv \le \unit[1200]{MeV}\). Given the strong dependence of
\thyp{} on \Esephyp{} and considering the large experimental
uncertainty of \Esephyp{}, none of the recent RHI reported \thyp{}
values listed in \Tab{tab:lifetime} can be excluded but, rather can be
correlated with its own underlying value of \Esephyp{}.
\begin{table}
  \centering
  \caption{Extrapolated \hyp{} \(\Lambda\) separation energies,
    two-body \(\pi^-\) decay rates, and lifetime values for several HO
    basis \UV{} scales. The
    \(E_{\mathrm{sep}}^{\mathrm{UV}}=E_\infty^{\mathrm{UV}} -
    E^{\mathrm{UV}}_{\nuc{2}{H}}\) is the extrapolated \(\Lambda\)
    separation energy calculated using the deuteron energy at the
    corresponding \luv{} scale. The uncertainty in \thyp{} comes from
    the experimental uncertainty in \(R_3\). The last column shows
    extrapolations to the STAR Collaboration~\cite{STAR:2019wjm}
    reported value of \Esephyp{}, see text for details. }
  \label{tab:uvdep}
  \begin{tabular}{l|llllll}
    \luv{}~(MeV) & 800 & 900 & 1000 & 1200 & 1400 & \(-\) \\ \hline
    \(E_{\mathrm{sep}}^{\mathrm{UV}}\)~(keV) &69&135&160&165&162&410 \\
    \(\Gamma_\infty^{\mathrm{UV}}\)~(GHz) &0.943&1.192&1.261&1.281&1.293&\starextrapg{} \\
    \thyp{}~(ps) &\(234\pm27\)&\(190\pm22\)&\(180\pm21\)&\(178\pm20\)&\(176\pm 20\)&\(\starextrapt{}\)
  \end{tabular}
\end{table}

Aside from the large measurement uncertainties, there are also
considerable theoretical uncertainties associated with \Esephyp{} and
\thyp{}. Note that the values listed in \Tab{tab:uvdep} are calculated
using a particular hypernuclear Hamiltonian. Recently, we explored the
limits of theoretical precision of
\Esephyp{}~\cite{Htun:2021jnu,Gazda:2022fte} and \Gtb{} due to model
uncertainties in \YN{} and \NN{}+\NNN{} interactions. For this purpose
we employed 4 versions of the \loyn{} interaction regularized at
\(\LamYN = 550, 600, 650, \unit[700]{MeV}\), together with the whole
family of 42 \nnlosim{} nuclear interactions
(see \Section{sec:yncsm}). Fixing \(\luv = \unit[1200]{MeV}\) for IR
extrapolations, the combined spread of \(\Lambda\) separation energies
for all combinations of \((\LamNN, \Tmax, \LamYN)\) is found to be
\(90 \lesssim \Esephyp \lesssim \unit[190]{keV}\), while the spread in
calculated two-body \(\pi^-\) rates
\(1.0 \lesssim \Gtb \lesssim \unit[1.4]{GHz}\), which implies
\(160 \lesssim \thyp \lesssim \unit[220]{ps}\) for a fixed value of
\(R_3 = 0.35\). Full details of this study will be presented in our
future work.

\section{Summary}
\label{sec:summary}
We performed a new microscopic calculation of the hypertriton
\(\pi^-\) two-body decay rate \Gtb{} employing \hyp{} and \he{}
three-body wave functions generated by \abi{} \YNCSM{} using realistic
\YN{} and \NN{}+\NNN{} interactions derived from \chieft{}. Using the
\(\Delta I=\frac12\) rule and the measured branching ratio \(R_3\) to
include the remaining \(\pi^0\) and 3-{} plus 4-body \hyp{} decay
channels, we deduced the \hyp{} lifetime \thyp{}. The main findings of
this study are:
(i) Replacing pionic \pw{} by realistic \(\pi^- {-} \he\) \dw{}
    enhances \Gtb{} by \(\approx 15\%\).
(ii) The \(\Sigma NN\) admixtures in \hyp{} reduce the purely
    \(\Lambda NN\) decay rate by \(\approx 10\%\) due to interference
    effects, despite their \(\lesssim 0.5\%\) contribution to the norm of
    the \hyp{} wave function.
(iii) The lifetime \thyp{} varies strongly with the rather poorly
    known \Esephyp{}. It is then possible to associate each of the
    distinct RHI measured \thyp{} values with its own underlying value of
    \Esephyp{}.
Remarkably, the most recent ALICE Collaboration's~\cite{ALICE:2022rib}
\(\Esephyp{} = \unit[72(63)(36)]{keV}\) central value is almost the
same as our lowest \(E_{\mathrm{sep}}^{\luv = 800}=\unit[69]{keV}\)
and our lifetime \(\tau^{\luv = 800}(\hyp)=\unit[234\pm 27]{ps}\) is
then consistent with their reported \(\thyp{}=253\unit[\pm 11 \pm
6]{ps}\). Only future experiments expected at MAMI, JLab, J-PARC, and
CERN will hopefully pin down \Esephyp{} with a better precision than
\unit[50]{keV} and lead to a resolution of the \hyp{} lifetime puzzle.

\section*{Acknowledgments}
We are grateful to Petr Navr\'{a}til for helpful advice on extending
the nuclear \NCSM{} codes to hypernuclei; to Johann Haidenbauer and
Andreas Nogga for providing us with the input \LO{} Bonn--J\"{u}lich
$\YN$ potential; and to Andreas Ekstr\"{o}m for providing us the
\nnlosim{} \NN{}+\NNN{} interactions used in the present work. The
work of D.G.\ was supported by the Czech Science Foundation GA\v{C}R
grants 19-19640S, 22-14497S, and by the European Union’s Horizon 2020
research and innovation program under grant agreement No~824093.
Some of the computational resources were supplied by IT4Innovations
Czech National Supercomputing Center supported by the Ministry of
Education, Youth and Sports of the Czech Republic through the
e-INFRA~CZ (ID:90140).
\bibliography{gpogf_hyp2022}
\end{document}